\documentclass[10pt]{article}

\usepackage[OE]{express}

\usepackage[utf8]{inputenc}
\usepackage{siunitx}             
\usepackage{graphicx}            
\usepackage{xcolor}   
\usepackage{amsmath}
\usepackage{amssymb}
\usepackage{mathtools}
\usepackage{braket}
\usepackage[version=3]{mhchem}
\usepackage{float}
\usepackage{hyperref} 
\usepackage{siunitx}
\hypersetup{colorlinks=true, citecolor=blue, colorlinks, linkcolor=blue, urlcolor=blue}

\begin{document}

\title{A simple, narrow, and robust atomic frequency reference at 993 nm exploiting the rubidium (Rb) $5\mathit{S}_{1/2}$ to $6\mathit{S}_{1/2}$ transition using one-color two-photon excitation }

\author{Thomas Nieddu,\authormark{1} Tridib Ray,\authormark{1} Krishnapriya S. Rajasree,\authormark{1} Ritayan Roy,\authormark{2,*} and S\'{i}le Nic Chormaic \authormark{1, 3, 4}}

\address{\authormark{1}Light-Matter Interactions Unit, Okinawa Institute of Science and Technology Graduate University, Onna, Okinawa 904-0495, Japan\\
\authormark{2}Quantum Systems \& Devices Group, School of Mathematical and Physical Sciences, Pevensey II, University of Sussex, Falmer, Brighton, BN1 9QH, UK\\
\authormark{3}Université Grenoble Alpes, CNRS, Grenoble INP, Institut Néel, 38000 Grenoble, France\\
\authormark{4}School of Chemistry and Physics, University of KwaZulu-Natal, Durban 4001, South Africa
}

\email{\authormark{*}ritayan.roy@u.nus.edu}

\begin{abstract}
We experimentally demonstrate a one-color two-photon transition from the $5\mathit{S}_{1/2}$ ground state to the $6\mathit{S}_{1/2}$ excited state in rubidium (Rb) vapor using a continuous wave laser at \SI{993}{nm}. The Rb vapor contains both isotopes ($^{85}$Rb and $^{87}$Rb) in their natural abundances. The electric dipole allowed transitions are characterized by varying the  power and polarization of the excitation laser.  Since the optical setup is relatively simple, and the energies of the allowed levels are impervious to  stray magnetic fields, this is an attractive choice for a frequency reference at \SI{993}{nm}, with possible applications in precision measurements and quantum information processing. 
\end{abstract}


\section{Introduction}
\label{sec:Introduction}

Two-photon processes in atomic systems have several distinct advantages over single-photon processes. For example, two-photon processes can give direct access to optical excitations that would be electric-dipole forbidden for a single-photon process \cite{shimoda2014}. Furthermore, when the two photons are derived from two laser beams in a counter-propagating configuration, a judicious choice of the polarizations  can yield background-less, Doppler-free spectra \cite{Vasilenko1970, Cagnac1973}. In addition, two-photon transition frequencies for  $\mathit{S}$ to $\mathit{S}$ transitions are insensitive to magnetic fields below the Pacshen-Back domain\cite{Bloembergen1974}, while two-photon transitions to metastable states have extremely narrow linewidths compared to those for single-photon processes \cite{Hansch2006,Blatt2003}. These unique features make two-photon spectroscopy a powerful tool for precision measurements. Following the first observation of a two-photon transition in an atomic system containing cesium \cite{Abella1962}, numerous different atomic transitions have been investigated \cite{Biraben1974, Roberts1975, Campani1978, YiWei2001, Bushaw2003, MingSheng2004, YiChi2010JphysB}. The technique has been extensively used for metrology and the accurate determination of fundamental constants \cite{Hansch1975,Matveev2013}, as a frequency reference \cite{Roy2017}, and in quantum telecommunications \cite{Kuzmich2006}.

In this paper, we report on the observation and spectroscopic study of the $5\mathit{S}_{1/2}$ to $6\mathit{S}_{1/2}$ two-photon transition in a hot Rb-atom vapor using a single-frequency laser beam. To our knowledge, this is the first observation of this particular Rb transition using a one-color two-photon excitation. We explore  the dependency of the spectroscopy signal on (i) the intensity and (ii) the polarization of the pump beam. Observation of a quadratic dependency on the intensity of the pump laser is a signature of the two-photon transition.  We also show that the pump laser frequency can be stabilized to the observed spectroscopic peaks, thereby illustrating that the transition can be used as a frequency reference.  Finally, we discuss some possible applications for precision measurements and  quantum telecommunications.

\section{Experimental Details}
\label{sec:Setup}

\begin{figure} [ht]
\centering
      \includegraphics[width=0.7\textwidth]{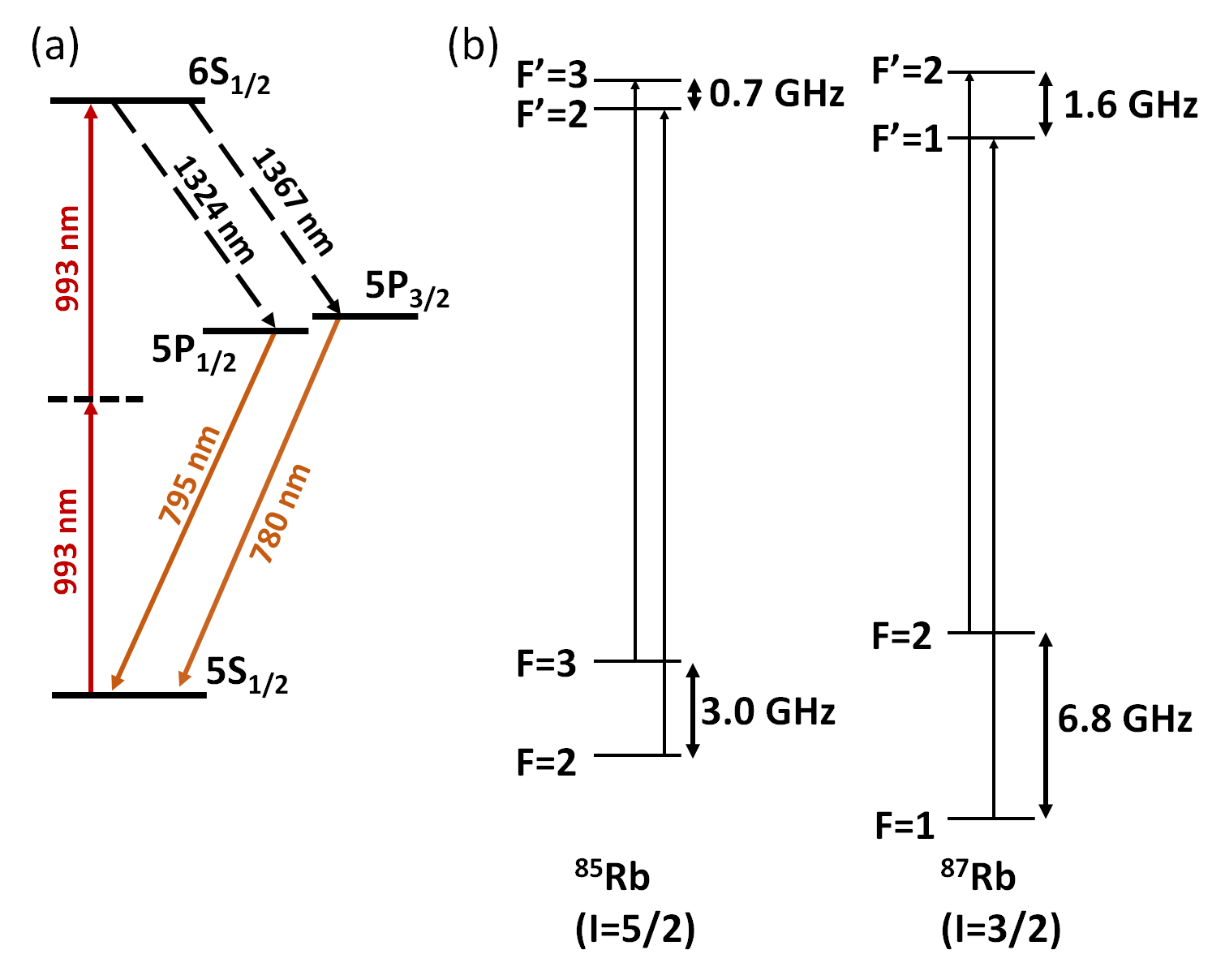}
  \caption{(a) Energy level diagram for Rb. A beam at \SI{993}{nm} excites atoms from $5\mathit{S}_{1/2}$ to $6\mathit{S}_{1/2}$ via single-color two-photon excitation. The intermediate virtual state is represented as a dashed line. The  atoms   decay back to $5\mathit{S}_{1/2}$ via   $5\mathit{P}_{1/2}$ or  $5\mathit{P}_{3/2}$, with  photons emitted   at 795 nm and 780 nm (orange arrows); (b) Hyperfine level diagrams for the two Rb isotopes. Two-photon transitions allowed by the selection rule, $\Delta F = 0$, $\Delta m_F = 0$, are shown, along with the frequencies of the hyperfine splittings.}
  \label{fig:Elevels}
\end{figure}

 The energy level diagram for the Rb transition of interest is shown in Fig. \ref{fig:Elevels}.  Atoms are excited from the $5\mathit{S}_{1/2}$ ground state via a virtual state to the $6\mathit{S}_{1/2}$  state using a two-photon process at \SI{993}{nm}.  The atoms can decay back to the ground state via two possible channels characterized by an intermediate state, which can be either (i)  the $5\mathit{P}_{1/2}$ level on emission of a pair of photons with wavelengths of \SI{1324}{nm} and \SI{795}{nm} (i.e. the D1 transition), or (ii)  the $5\mathit{P}_{3/2}$ level on  emission of a pair of photons with wavelengths of \SI{1367}{nm} and \SI{780}{nm} (i.e. the D2 transition).  The photons at \SI{1324}{nm} and \SI{1367}{nm} fall beyond the range of the detectors available for this experiment, hence, for the   work reported hereafter, we  only detect the \SI{780}{nm} and \SI{795}{nm} light.   
 
\begin{figure} [ht]
\centering
\includegraphics[width=10 cm]{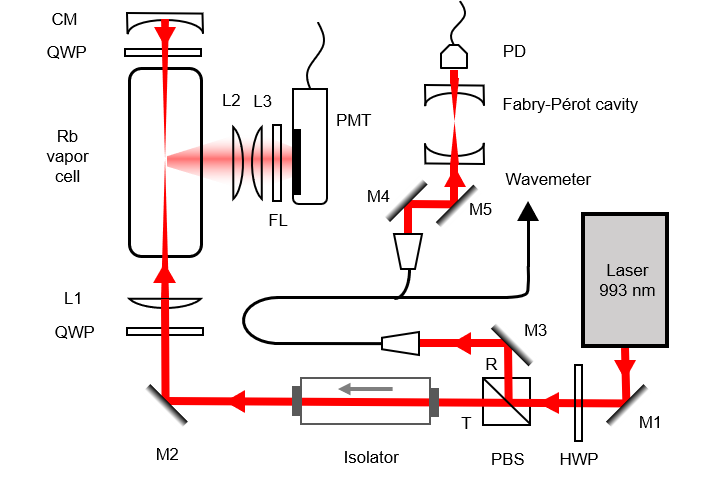}
  \caption{Schematic  of the experimental setup. Light from a tunable 993 nm laser is used for  two-photon excitation in a Rb vapor cell using a retro-reflected configuration.  The resulting atomic fluorescence is detected by a PMT. The polarizations of the forward and retro-reflected beams are controlled using QWPs.  A small amount (i.e. <10\%) of the  993 nm beam is coupled to a Fabry-P\'erot cavity and a wavemeter to monitor the laser frequency. M1-M5: Mirrors, L1-L3 Plano-convex lens, HWP: Half-wave plate, QWP: Quarter-wave plate, PBS: Polarizing beam splitter, CM: Concave mirror, PMT: Photomultiplier tube, FL: Short-pass optical filter, PD: Photodiode.} 
  
  \label{fig:setup}
\end{figure}

 The experimental setup is illustrated in Fig. \ref{fig:setup}. The experiment is carried out in a glass cell filled with Rb in its natural isotopic abundances, maintained at a temperature of \SI{130}{\celsius}. The \SI{993}{nm} beam used to drive the two-photon transition is provided by a continuous wave (CW) Ti:Sapphire laser (Coherent MBR 110), locked to a scanning reference cavity yielding a spectral linewidth of \SI{100}{kHz}. The laser frequency can be scanned by changing the length of the reference cavity. The combination of a half-wave plate (HWP) and a polarizing beam-splitter (PBS) at the output of the laser allows us to control the powers in the reflected (R) and transmitted (T) beams from the PBS. Most of the optical power, typically >90\%, is in T and passes through the vapor cell for the two-photon spectroscopy studies. The remaining light, in R, is fiber-coupled and further split so that 99\% goes to a Fabry-P\'erot cavity (Toptica FPI-100) and 1\%  to a wavemeter (HighFinesse WS-6). The wavemeter has two purposes; it allows us to tune the laser to the desired wavelength and to monitor the frequency scanning. The Fabry-P\'erot cavity has a free-spectral range of \SI{1}{GHz} and is used to monitor the linearity of the frequency scan.

 An optical isolator is placed before the vapor cell to avoid reflections back into the laser. A plano-convex lens (L1), with focal length $f_1 = \SI{150} {\mm}$, is placed after the optical isolator to focus the beam in the cell. The $1/e^2$ beam diameter is \SI{128}{\mu m}. A concave mirror (CM), with focal length $f_{CM} = \SI{75} {\mm}$, and placed $2f_{CM} = \SI{150} {\mm}$ away from the focal plane of L1 ensures retro-reflection of the beam back to the focal point. Quarter-wave plates (QWP) can be inserted in the beam path before L1 and  CM to generate a circularly polarized beam.
We detect both the \SI{795}{nm} and \SI{780}{nm} decay photons using a photomultiplier tube (PMT) (Hamamatsu R636-10).  A short-pass filter with a cut-off wavelength of \SI{800}{nm} is placed in front of the PMT to prevent any scattered light from the \SI{993}{nm} pump from being detected. We use a pair of lenses, L2 and L3, with focal lengths of $f_2 = f_3 = \SI{50} {\mm}$, in front of the PMT  in a telescope configuration for efficient collection of the light. The quantum efficiencies of the PMT at \SI{780}{nm} and \SI{795}{nm} are 9\% and 8\%, respectively. The obtained current is amplified by a pre-amplifier with a gain of $10^5$ and dropped across a 50 $\Omega$ resistor on an oscilloscope.

\section{Results}
\label{sec:Spectroscopy}
\begin{figure} 
\centering
      \includegraphics[width=\textwidth]{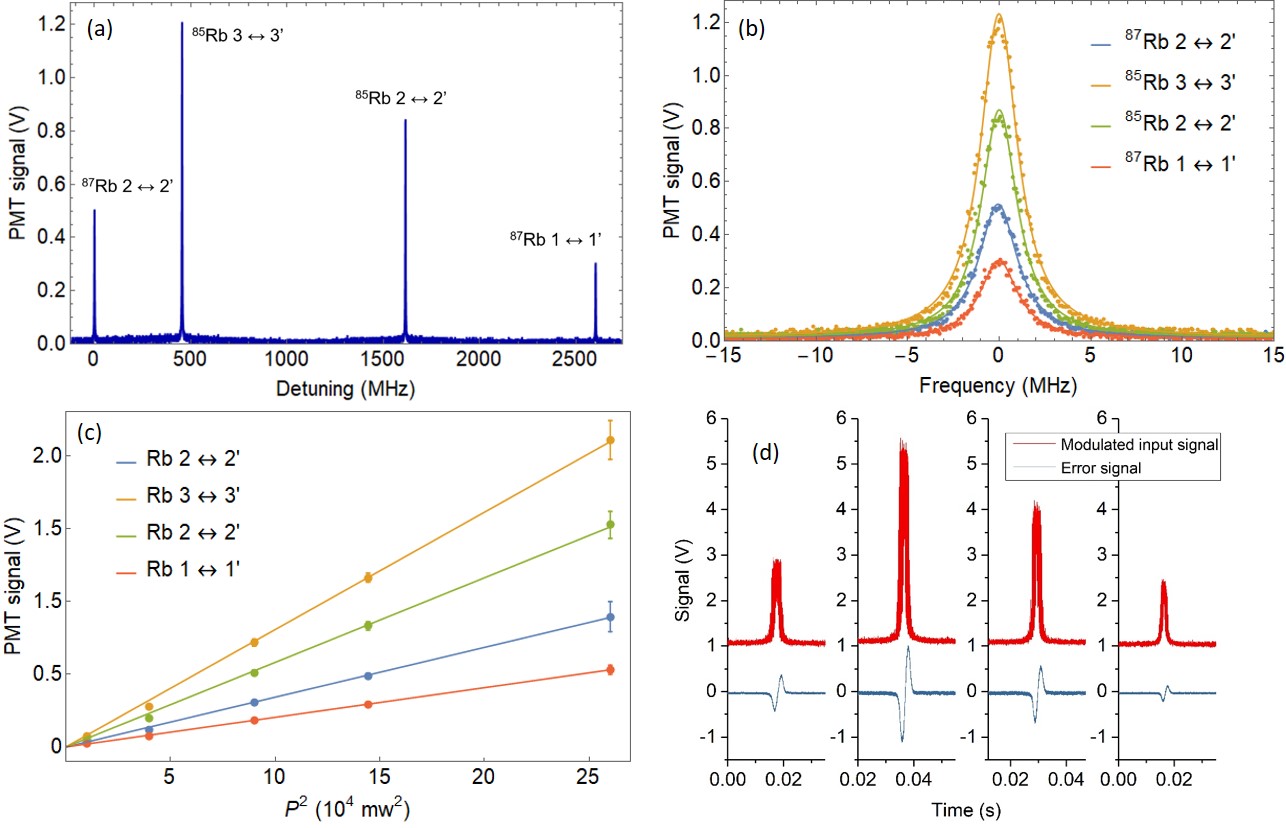}
  \caption{(a) Typical spectroscopic signal obtained by scanning the frequency of the \SI{993}{nm} pump beam and recording the signal on the PMT. Each peak indicates a hyperfine transition as labeled. (b) Comparison of the individual peak intensities and linewidths from (a). (c) Linear dependence of the peak height as a function of the total pump  power (P) squared. (d) Modulated signals and the generated error signals for each peak to which the laser can be locked. For clarity, a 1 V offset is added to the modulated signal.}
  \label{fig:ExptRes}
\end{figure}

Electric dipole allowed two-photon transitions from $\mathit{S}$ to $\mathit{S}$ levels obey the selection rules $\Delta F = 0$ and $\Delta m_F = 0$ \cite{Bonin1984}. This results in only two allowed transitions for each Rb isotope, i.e. $^{87}$Rb $F=2 - F'=2$, $^{85}$Rb $F=3 - F'=3$, $^{85}$Rb $F=2 - F'=2$, and $^{87}$Rb $F=1 - F'=1$ (see Fig. \ref{fig:Elevels}(b)). A typical spectrum obtained is shown in Fig. \ref{fig:ExptRes}(a).  Here,  excitation to the $6\mathit{S}_{1/2}$ level is obtained by scanning the frequency of the 993 nm laser and using the same linear polarization for the forward and  retro-reflected beams that generate the two-photon process.  Note that the simple setup presented here does not measure the absolute frequency of the transition. The hyperfine splitting of the $6\mathit{S}_{1/2}$ level is obtained from  Ref.\cite{Galvan2008} using a resonant intermediate level. The relative frequency difference is obtained by setting the  frequency of the first peak, i.e. the $^{87}$Rb $5\mathit{S}_{1/2}, F=2$ - $6\mathit{S}_{1/2}, F'=2$ transition, to zero, see Fig. \ref{fig:ExptRes}(a). A linear frequency scaling obtained by using fringe interpolation of the Fabry-P\'erot peaks, which are 1 GHz apart, yields a similar result. Due to the two-photon process, the relative frequency differences of the peaks are  half the  actual energy differences of the atomic levels. 

\subsection{Power variation}

The relative height and width of each peak in Fig. \ref{fig:ExptRes}(a) is shown in Fig. \ref{fig:ExptRes}(b). For a particular Rb isotope, the intensities of the transitions from the ground hyperfine levels are proportional to the statistical weights of the atomic population in those hyperfine levels \cite{Grynberg1977b}. However, since the difference in energy between the hyperfine levels is negligible compared to the transition energy, the weight factor  is equivalent to  the degeneracy $(2F+1)$ of the hyperfine levels. These values are $5:3$ and $7:5$ for $^{87}$Rb and $^{85}$Rb, respectively. The variation of  peak height as a function of the square of the laser power is shown in Fig. \ref{fig:ExptRes}(c). The peak heights show a quadratic dependence on the total  beam power, $P$ (i.e. the sum of the powers in the forward and retro-reflected beams); this is a signature of the two-photon process \cite{Grynberg1977}. The ratios of the slopes of the fitted straight lines are 1.667 for  $^{87}$Rb and 1.387  for $^{85}$Rb, i.e., close to the expected ratios of $5:3$ and $7:5$, respectively. The width of the peaks does not change as a function of power, at least within the standard deviation of the measurements. We measure a Lorentzian full-width-at-half-maximum (FWHM) of $2.60\pm 0.07$ MHz, $2.44\pm 0.09$ MHz, $2.49\pm 0.04$ MHz and $2.43\pm 0.04$ MHz for $^{87}$Rb $F=2 - F'=2$,$^{87}$Rb $F=1 - F'=1$, $^{85}$Rb $F=3 - F'=3$ and $^{85}$Rb $F=2 - F'=2$ peaks, respectively.  

\subsection{Laser frequency stabilization to spectroscopic peaks }
To establish the viability of the transition as a frequency reference, we demonstrate frequency locking of the pump laser to the spectroscopic peaks. This is implemented by integrating a TEM LaseLock\textsuperscript{\textregistered}  module with the Ti:Sapphire laser.  First, the reference cavity is bypassed and the laser frequency is scanned by directly scanning  its cavity. A 10 kHz  modulation is applied  to one of the piezo-driven mirrors to generate frequency sidebands. The modulated spectroscopic signal is fed into a lock-in amplifier and an error signal is generated. An example of a modulated signal and a derived error signal are shown in Fig. \ref{fig:ExptRes}(d). The laser cavity and, hence, the frequency can be stabilized to each of these error signals. 

\subsection{Polarization variation}

We next explore the effect of changing the polarization of the counter-propagating beams on the spectroscopic signal. As shown in Fig. \ref{fig:setup}, a quarter-wave plate  at either end of the vapor cell can be used to generate identical, or orthogonal, circularly polarized forward and retro-reflected beams. Let us denote linear and circular polarizations as $\pi$ and $\sigma$, respectively, and their orthogonal polarizations as $\pi'$ and $\sigma'$, respectively. As  mentioned, the two-photon transition selection rules between the $S$ levels are $\Delta F = 0$ and $\Delta m_F = 0$, hence the total angular momentum of  two photons absorbed by an atom during the excitation process must be zero.

\begin{figure} [ht]
\centering
      \includegraphics[width=\textwidth]{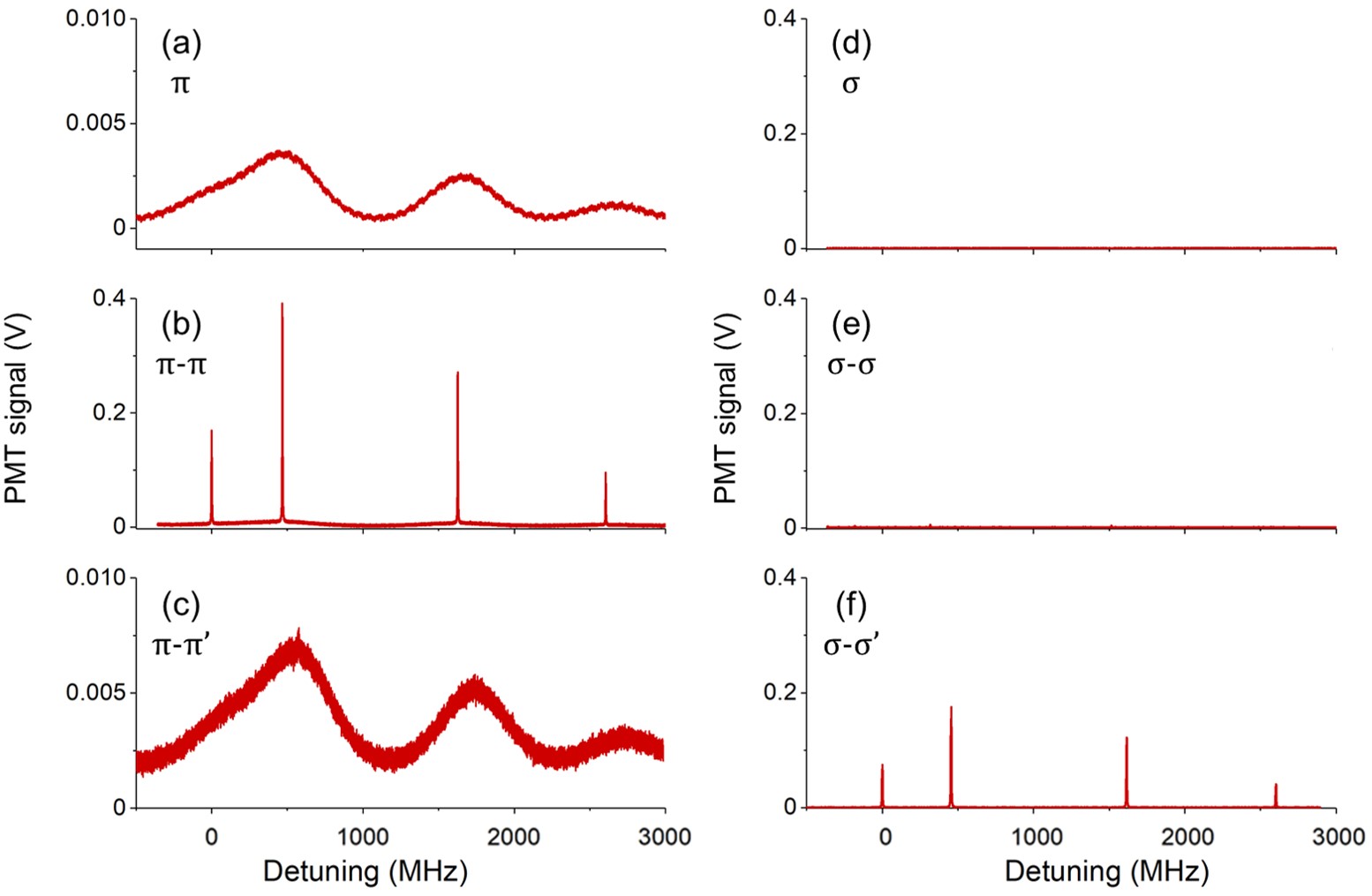}
  \caption{Effect of beam polarization on the two-photon excitation as recorded by the PMT. As in Fig. \ref{fig:ExptRes}(a), the relative frequency  is obtained by setting the  frequency of the $5\mathit{S}_{1/2}, F=2$ - $6\mathit{S}_{1/2}, F'=2$ peak to zero. The power of the 993 nm beam is fixed at 250 mW and its frequency is scanned. The polarization of the beam is changed using QWPs. (a) Doppler-broadened spectrum with a single, linearly polarized beam. (b) Linearly polarized counter-propagating beams reveal the Doppler-free peaks and a small Doppler-broadened base. (c) Counter-propagating beams with orthogonal linear polarizations yield a Doppler-broadened spectrum of twice the amplitude of that in (a). (d) A single, circularly polarized beam does not yield a signal. This configuration is forbidden, according to the selection rules. (e) Counter-propagating beams with identical circular polarizations do not yield a signal for the same reason as in (d). (f) Counter-propagating beams with orthogonal circular polarizations yield a background-less Doppler-free spectrum.  Here, the total angular momentum for the transition is zero.} 
  \label{fig:pola}
\end{figure}

First we study the case of linear polarizations. By blocking the retro-reflected beam, a $\pi$ configuration is created, see Fig. \ref{fig:pola}(a).  We observe a Doppler-broadened signal  since both  photons are derived from the forward beam. Next, by introducing a retro-reflected beam, a $\pi\,-\,\pi$ configuration is created, see Fig. \ref{fig:pola}(b). Here, we obtain a narrow Doppler-free spectrum on top of a small Doppler-broadened baseline. The Doppler-free spectrum arises when the two photons are absorbed from counter-propagating beams, whereas the Doppler-broadened signal results from the two photons being absorbed from the same (forward or retro-reflected) beam.  Next, the addition of a QWP after the vapor cell, aligned at \ang{45}  with respect to the forward beam's polarization axis,  creates a linearly polarized, retro-reflected beam, orthogonal to the forward beam, resulting in a $\pi\,-\,\pi'$ configuration, see Fig. \ref{fig:pola}(c). In this case, the two photons can only be absorbed from either the forward or the retro-reflected beam, i.e., they cannot be absorbed simultaneously from both beams. This results in a signal on the PMT that is the sum of two Doppler-broadened spectra, one from each beam, yielding double the amplitude of the $\pi$ configuration. 

We next move to the case where the beams have circular polarizations. By inserting a QWP at \ang{45} before the vapor cell and blocking the retro-reflected beam, a $\sigma$ configuration (see Fig. \ref{fig:pola}(d)) is created. In this case, the transition is forbidden (the sum of the angular momenta of two photons in the forward beam is non-zero), hence there is no signal recorded on the PMT. The $\sigma\,-\,\sigma$ configuration (see Fig. \ref{fig:pola}(e)) is created  by allowing the retro-reflected beam to propagate inside the vapor cell. This transition is also forbidden as, once more, the sum of the angular momenta of the two photons from the  counter-propagating beams ($2\hbar$) is  non-zero; as a result there is no signal. Finally, inserting a QWP before the retro-reflecting mirror  forms a $\sigma\,-\,\sigma'$ configuration. The orientation of the waveplate's axis is irrelevant. The two photons that drive the transition can only be absorbed from the counter-propagating beams. As a result, a background-less, Doppler-free spectrum is obtained (Fig. \ref{fig:pola}(f)). The peak heights are half those obtained for the  $\pi\,-\,\pi$ configuration as the probability to absorb two-photons with opposite spin angular momentum is lesser in this case. It should be noted that the other orthogonal circular polarization configuration yields a similar result and is not presented here.

\section{Discussion}
\label{sec:Discussion}

\textit{Ab-initio }calculations of the electronic wavefunction close to the nucleus rely on an accurate measurement of the hyperfine splitting of the atomic energy level \cite{Gomez2004}. To date, the $6\mathit{S}_{1/2}$ level in Rb has been accessed via a two-color, two-photon excitation scheme at  \SI{795}{nm} and \SI{1324}{nm}  for measuring its lifetime \cite{Galvan2008} and hyperfine splitting \cite{Gomez2005}. Accessing the $6\mathit{S}_{1/2}$ level via the one-color, two-photon method presented here should enhance the accuracy and precision of such measurements since only a single laser is necessary for the excitation. Additionally, the two-photon transition could be used for the measurement of parity nonconservation \cite{wieman99,Sarkisyan2003} in alkali atoms.  

The excitation scheme presented herein enables the conversion of two near-infrared photons at \SI{993}{nm} into a telecommunication O-band photon, at either \SI{1324}{nm} or \SI{1367}{nm}, and another near-infrared photon, at \SI{795}{nm} or \SI{780}{nm}, respectively. Chaneli\`ere \textit{et al.} \cite{Kuzmich2006} proposed a method for building quantum repeaters  using  cascaded atomic transitions, whereas Willis \textit{et al.} \cite{Willis2008}  generated time-correlated photon-pairs between a near-infrared photon and an  O-band photon using a four-wave mixing (4WM) process in a  Rb vapor. The 4WM scheme made use of the $6\mathit{S}_{1/2}$ level, accessed via a two-color, two-photon excitation. 
Our scheme is compatible with these results, as driving the $5\mathit{S}_{1/2}$ to $6\mathit{S}_{1/2}$ two-photon transition would permit us to exploit both O-band photons as signal photons, and the corresponding NIR photons as idlers mapped onto an atomic quantum memory. In particular, the \SI{993}{nm} photons could be coupled to atoms interacting with the evanescent field at the waist of an optical nanofiber embedded in a cold atomic ensemble \cite{Thomas2016} to make the process more efficient \cite{Franson2008}. 

The first \-order Zeeman shifts experienced by the same hyperfine states of the $5\mathit{S}_{1/2}$ and  $6\mathit{S}_{1/2}$ levels are identical since  they have the same hyperfine Land\'e g-factors. This feature renders the frequency of the $5\mathit{S}_{1/2}$ to  $6\mathit{S}_{1/2}$ transition insensitive to stray magnetic fields. The transition frequency is also less sensitive to electric fields compared to  transitions to nonzero angular momentum states (where $l>0$). These features makes the transition an attractive choice for a frequency reference. 

\section{Conclusion}
\label{sec:Conclusion}
We have demonstrated the  $5\mathit{S}_{1/2}$  to $6\mathit{S}_{1/2}$ one-color, two-photon transition in a hot Rb vapor. The effects of excitation laser power and beam polarization on the observed spectroscopy signals were investigated. We also propose that the transition can be used as a reference at 993 nm by demonstrating frequency stabilization of the excitation laser to the spectroscopic peaks. The simple optical setup is easy to miniaturize and can be readily integrated into more complex experiments. The transition frequency is insensitive to stray magnetic fields and is, therefore, suitable for precision measurements and experimental setups where magnetic fields cannot be completely eliminated, e.g. in a magneto-optical trap or a magnetically trapped Bose-Einstein condensate. Future investigation of the enhanced nonlinear process by embedding an optical nanofiber  in such a system will open up new possibilities for the generation of a fiber-integrated photon-pair source for quantum key distribution.

\section*{Funding}
This work is supported by the Okinawa Institute of Science and Technology Graduate University. 

\section*{Acknowledgments}
R. Roy would like to thank B.Hessmo for discussion.
\end{document}